\documentclass[letter]{aa}  

\usepackage{natbib}
\usepackage{graphicx}
\usepackage{amsmath}
\usepackage[varg]{txfonts}

\begin{document}

\title{TH\'EMIS: a calibration-free solar telescope}

   
\author{V. Bommier\inst{1}
        \and B. Gelly\inst{2}
        \and R. Douet\inst{2}}
       
\institute{LIRA, Observatoire de Paris, Universit\'e PSL, CNRS, Sorbonne Universit\'e,
                  Universit\'e Paris Cit\'e, 5 place Jules Janssen, F-92195 Meudon, France\\
             \email{V.Bommier@observatoiredeparis.psl.eu}
            \and TH\'EMIS, Tenerife, Canary Islands, Spain\\ }

\date{Received ... / Accepted ...}

 
\abstract
{TH\'EMIS was initially built as a calibration-free telescope, with the polarisation analysis performed in F1 focus, before any oblique reflection. Recently, the telescope was improved with installation of an Adaptive Optics apparatus. }
{We would like to show you that it remains a calibration-free telescope, even if the two beams able to provide the light polarisation, are now separated in front of the cameras only.}
{This is obtained owing to an adaptation of all the oblique reflections along the ray path.}
{We present raw images taken across a sunspot, with three lines: two polarisable lines, which display the sunspot magnetic field polarisation, linear and circular, and one unpolarisable line, which does not display any polarisation, which shows that the telescope is essentially free of any instrumental polarisation, because, if this line had showed any polarisation, this polarisation would have been of instrumental origin. Those images were recorded on September 12, 2024. We also present vector magnetic field, vector current density and vector Lorentz force maps obtained in July 2025, which are in accordance with previous results: circular currents wrapping spots clockwise about a positive polarity spot and counterclockwise about a negative polarity spot, and centripetal Lorentz force maintaining sunspots.}
{As a result, TH\'EMIS remains a calibration-free telescope.}
 
\keywords{Techniques: polarimetric -- Sun: magnetic fields}

\maketitle

\section{Introduction}

The French T\'elescope H\'eliographique pour l'\'Etude du Magn\'etisme et des Instabilit\'es Solaires (TH\'EMIS) is implemented in the European Iza\~na site located on the Tenerife Island (Canary Islands, Spain) \citep[see the review by][]{Schmieder-etal-25}. The telescope itself stands on top of a 25 m height tower and features a primary mirror 92 cm in diameter. It has been operating since 1998. It is devoted to the full polarisation mapping of the Sun's surface, for magnetic field measurements. In this respect, it was conceived as free of any instrumental polarisation, so that polarisation observation results can be derived without any need of polarisation calibration. It was then conceived as a calibration-free telescope. The method for this lies in the polarisation analysis performed in the F1 focus, before any oblique reflection. This is detailed in Sect. \ref{before2016}.

However, the telescope was modernised in 2015-2021 with introduction of Adaptive Optics \citep{Gelly-etal-16}. This implied a thorough modification of the light path inside the telescope, and consequently of the polarisation analysis, although always achieved in F1 focus. Adaptive Optics required to reduce the previous double beam for polarimetry, into a single beam coming on the adaptive mirror. This adaptation is described in Sect. \ref{after2016}. This involves not only the F1 focus, but also all the oblique reflections hereafter in the telescope, which makes it now calibration-free up in front of the cameras, where the two beams for polarimetry are now finally separated. 

The aim of the present paper is to show that this telescope is effectively free of any instrumental polarisation, by reporting the polarisation observed in an unpolarisable line, Fe I 5576.1 \AA, which has a Land\'e factor equal to zero, so that observed polarisation in it would be of instrumental origin. These observations, which confirm the calibration-free character of the telescope as conclusion of the present paper, are reported in Sect. \ref{instpol}. In Sect. \ref{obs}, we present recent observation results in sunspots.

\section{Polarimetric Analysis before 2016}
\label{before2016}

The TH\'EMIS telescope was conceived with polarimetric analysis performed in the primary focus F1. The polarisation analyser in F1 split the beam into two beams of respective intensity $I + S$ and $I - S$, where $S$ is one of the Stokes parameters $Q$, $U$, or $V$, and $I$ is the initial beam intensity, in such a way that the researched Stokes parameter was finally derived by subtracting the two beam intensities, which were recorded by the cameras.

These two beams were propagating in the telescope and spectrograph in a parallel manner. However, this double beam structure was not compatible with introduction of an adaptive mirror, which requires the beam to be single.

\section{The 2016 modification for Adaptive Optics introduction}
\label{after2016}

The requirement for Adaptive Optics was then a single beam inside the telescope. The polarisation analyser, always placed in the polarisation-free focus F1, was then modified and complemented with an anti-separator. The dual beam principle is preserved, but now as a dual superimposed beam splitter. This splitter creates two superimposed beams, each beam being now identified by a distinctive polarisation state: one linearly polarised along a reference direction and the other along the perpendicular direction.

These beams propagate through the transfer optics, and are separated before the spectral camera, by using that linear polarisation property, which then assumes that the transfer optics will preserve the linear polarisation states. 

This was obtained by replacing each 45\textsuperscript{o} reflection by two 22.5\textsuperscript{o} reflections. This improves the conservation of the linear polarisation in the reflection, in a non-linear manner, because two 22.5\textsuperscript{o} reflections present much less retardance than one at 45\textsuperscript{o}. The retardance at incidence 22.5\textsuperscript{o} is considerably lower than in the 45\textsuperscript{o} case, with a variation range of $\pm 2.8\textsuperscript{o}$ compared to the $\pm 15\textsuperscript{o}$ at incidence 45\textsuperscript{o}.

This was secondly obtained by modifying the field derotator, which is compensating for the image rotation on the spectrograph entrance slit. This derotator was transformed into an innovative design made from two Crova prisms and one flat mirror, whose Mueller matrices compensate each other. Each Crova prism contains in fact two close 45\textsuperscript{o} incidence reflections, but almost crossed, at 100\textsuperscript{o}, which make them compensating for.

As a result, the linear polarisation of the beam is practically not modified along the beam path, and the two superimposed beams are now separated before the spectral camera, into two beams of respective intensity $I + S$ and $I - S$, where $S$ is the searched for Stokes parameter.

\section{Test of the instrumental polarization}
\label{instpol}

In this respect, the TH\'EMIS telescope is now calibration-free up in front of the cameras. In order to prove this property, we present below the polarisation observed in an unpolarisable spectral line, Fe I 5576.1 \AA, which was used in the past to test the instrumental polarisation \citep{Bommier-Rayrole-02}. This line belongs to multiplet 686 of Fe I. Its upper level is $L = 2$, $S = 2$ and $J=0$, which results in one single magnetic sublevel and therefore no Land\'e factor. Its lower level is $L = 3$, $S = 2$, $J=1$, which results into Land\'e factor $g = 0$. This line cannot then display any Zeeman or Hanle effect of solar origin. If this line would anyway display any polarisation, this polarisation would then be of instrumental origin.

\begin{figure}
\includegraphics[width=8.8cm]{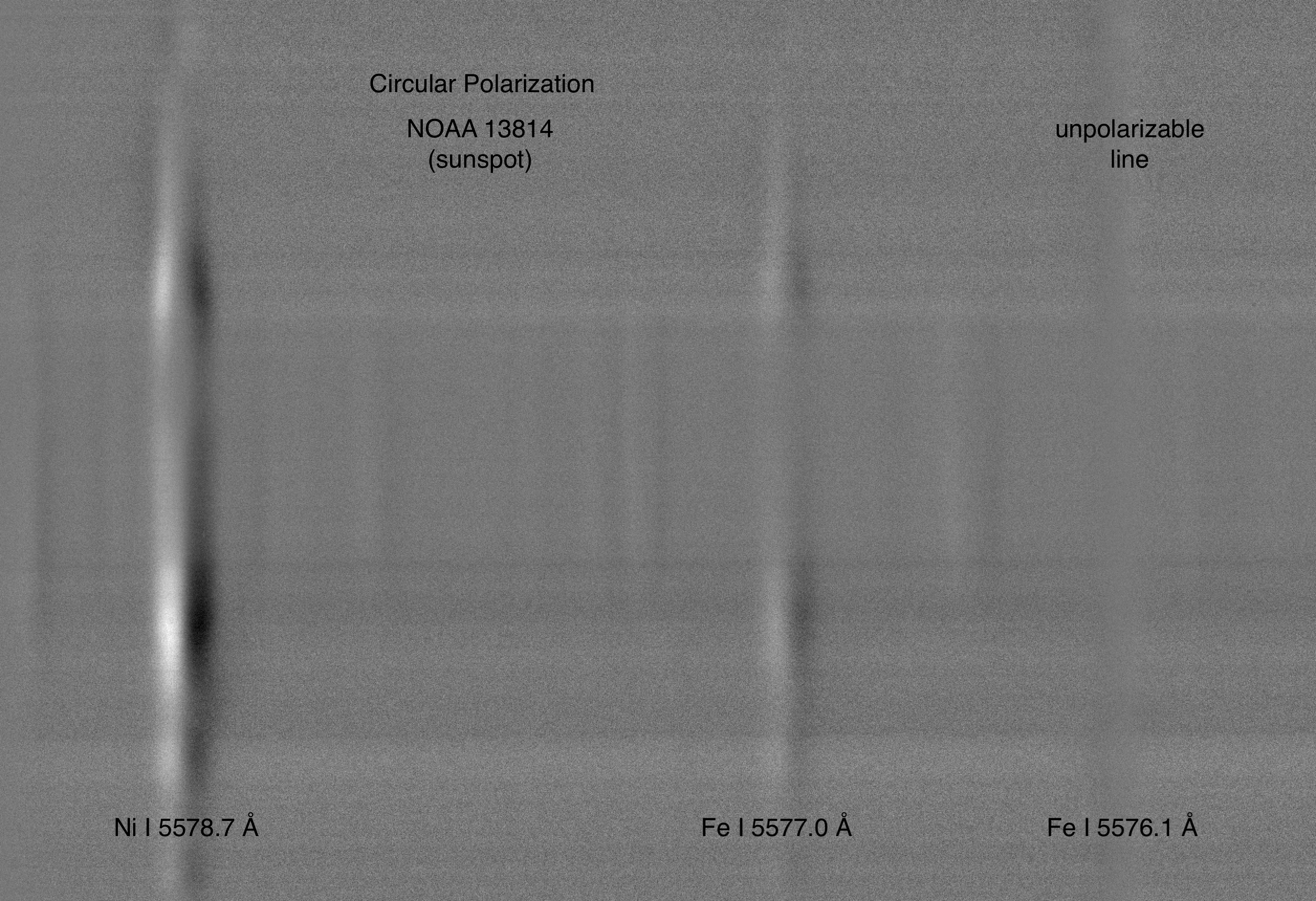}
\caption{Raw spectrum of Stokes $V$ (double value) across active region NOAA 13814. The two lines Ni I 5578.7 \AA\ and Fe I  5577.0 \AA\ are polarisable, when the third line Fe I 5576.1 \AA\ is unpolarisable.}
\label{2V}
\end{figure}

\begin{figure}
\includegraphics[width=8.8cm]{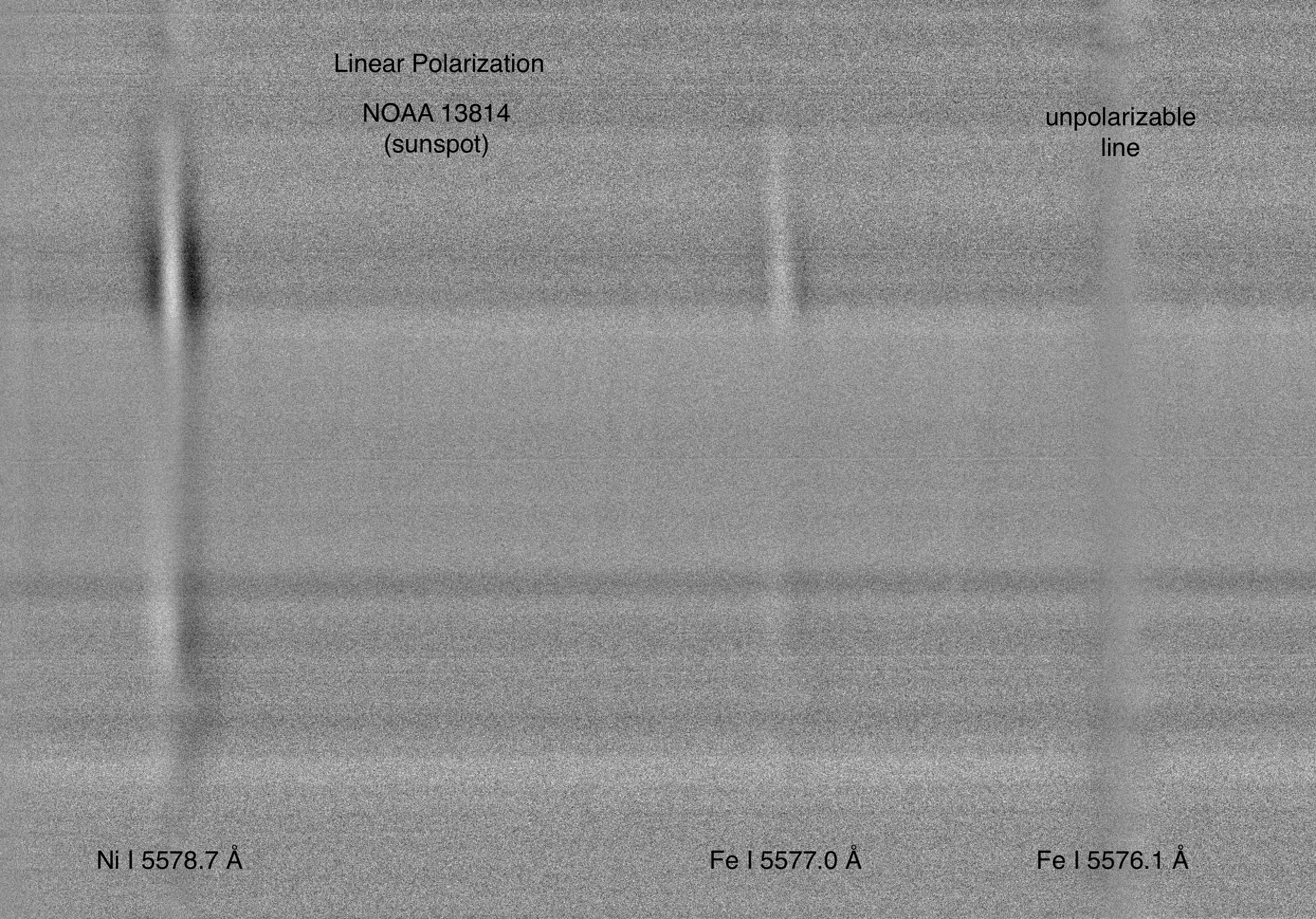}
\caption{Raw spectrum of Stokes $U$ (double value) across active region NOAA 13814. The two lines Ni I 5578.7 \AA\ and Fe I  5577.0 \AA\ are polarisable, when the third line Fe I 5576.1 \AA\ is unpolarisable.}
\label{2U}
\end{figure}

In Figs \ref{2V} and \ref{2U}, we present raw images subtraction providing circular and linear polarisation respectively. The reference axis for the linear polarisation is the local parallel. These images were observed across the sunspot of active region NOAA 13814 observed on September 12, 2024. No polarisation can be observed in the unpolarisable line, when the two other lines of the spectrum display solar polarisation. The absence of polarisation in the unpolarisable line confirms that the instrumental polarisation remains negligible in the TH\'EMIS telescope up in front of the cameras.

\section{Mueller matrix of the telescope}
\label{mueller}

The Mueller matrix $\vec{M}$ of the telescope at about 6000 \AA\ has been determined to be:
\begin{equation}
\vec{M} = \begin{pmatrix}
1.000 & -0.009 & -0.003 & 0.001 \\
-0.008 & 0.885 & 0.016 & -0.033 \\
0.014 & -0.436 & 0.872 & 0.033 \\
-0.019 & 0.415 & 0.008 & 0.873
\end{pmatrix} .
\end{equation}
It is averaged over one full day. It is quite constant along one day. It includes changing elevation axis and field derotation.

This matrix can be taken into account in the data treatment chain, by multiplying the output Stokes parameters by its inverse matrix $\vec{M^{-1}}$, which is:
\begin{equation}
\vec{M^{-1}} = \begin{pmatrix}
1.000 & 0.012 & 0.003 & 0.001 \\
0.010 & 1.100 & -0.021 & 0.042 \\
-0.012 & 0.570 & 1.136 & -0.021 \\
0.017 & -0.528 & -0.001 & 1.126
\end{pmatrix} .
\end{equation}
As a result, the Stokes parameters, which enter the telescope, are obtained. However, no significant change has been observed in the derived magnetic field maps, in several test cases. The inverse matrix is curiously very similar to the direct matrix.

As a result, the THEMIS telescope may be considered as calibration, and the inverse Mueller matrix may be applied for security.

\section{Vector magnetic field observations}
\label{obs}

\begin{figure}
\includegraphics[width=5.8cm]{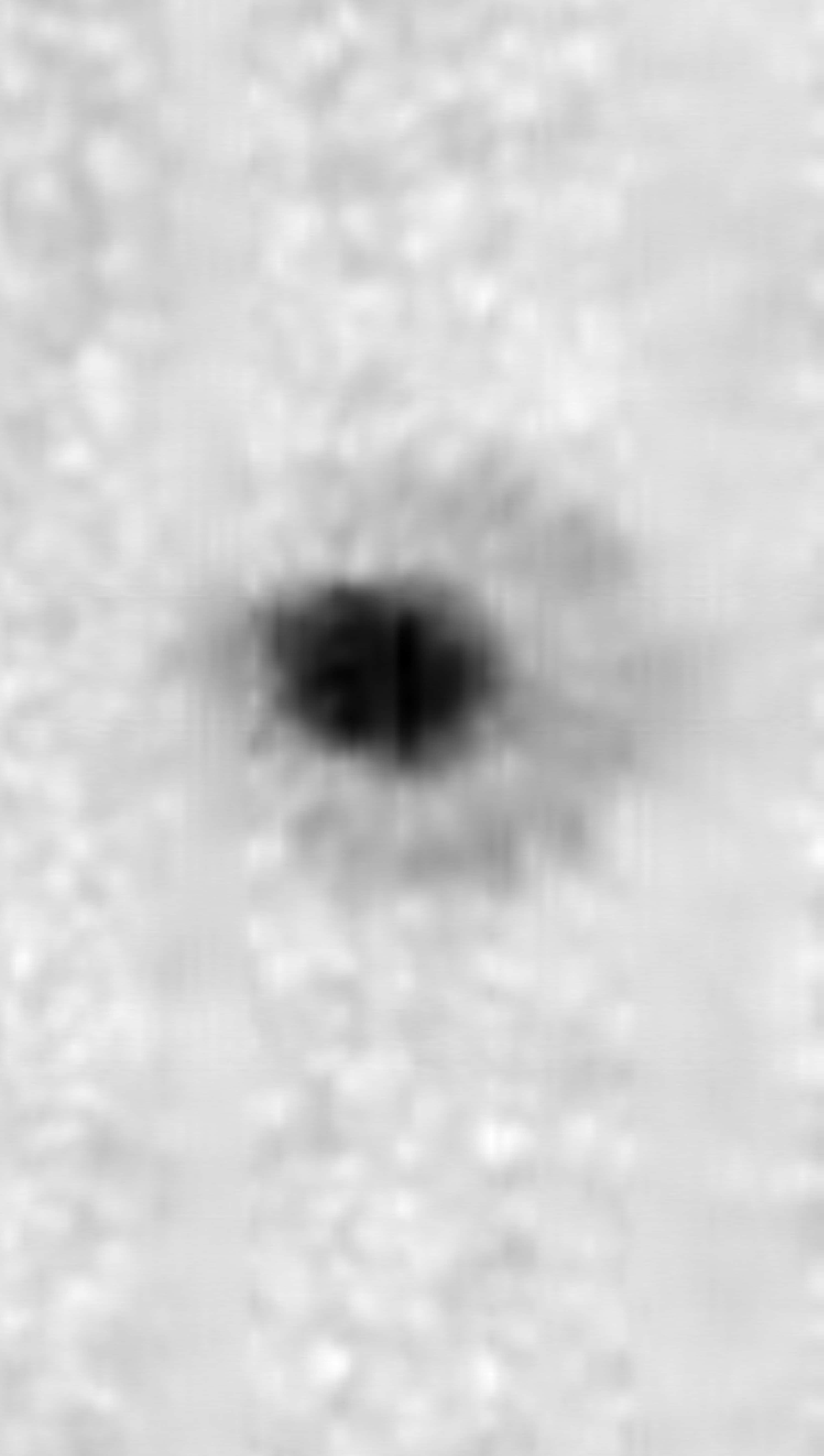}
\caption{West sunspot in NOAA 14142 observed on July 16, 2025.}
\label{I-16July}
\end{figure}

\begin{figure}
\includegraphics[width=5.8cm]{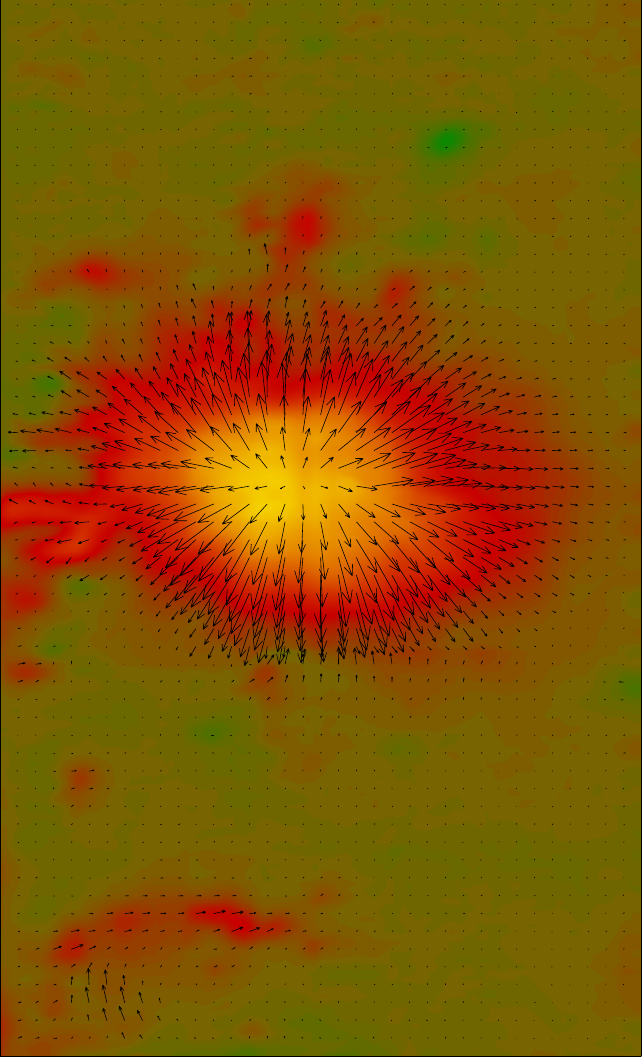}
\includegraphics[width=1.5cm]{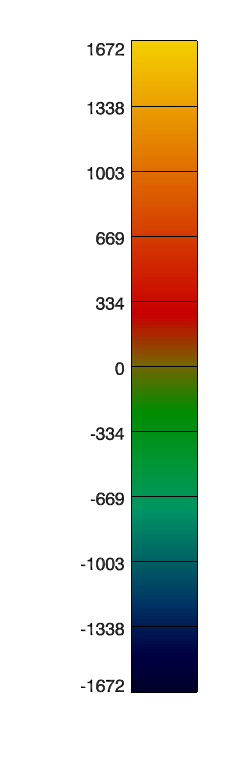}
\includegraphics[width=1.5cm]{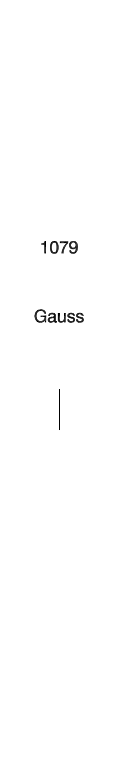}
\caption{West sunspot in NOAA 14142 observed on July 16, 2025: vector magnetic field.}
\label{B-16July}
\end{figure}

\begin{figure}
\includegraphics[width=5.8cm]{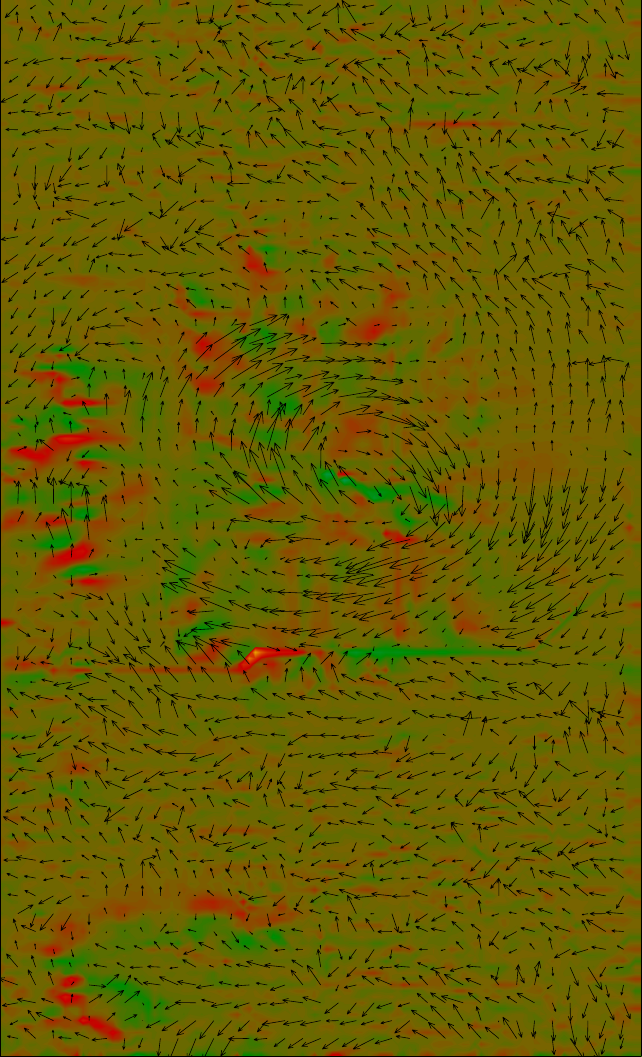}
\includegraphics[width=1.5cm]{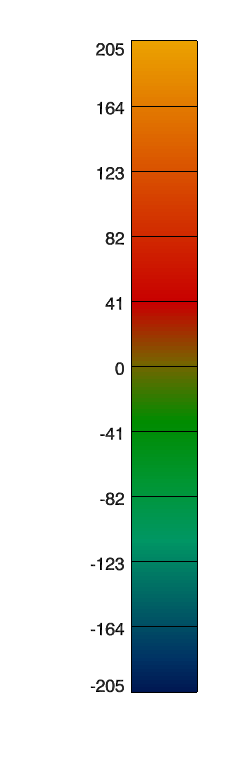}
\includegraphics[width=1.5cm]{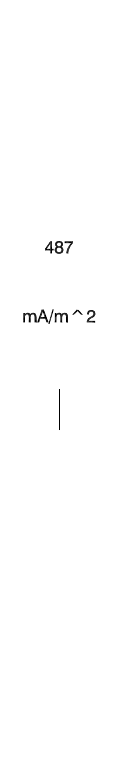}
\caption{West sunspot in NOAA 14142 observed on July 16, 2025: vector current density.}
\label{J-16July}
\end{figure}

\begin{figure}
\includegraphics[width=5.8cm]{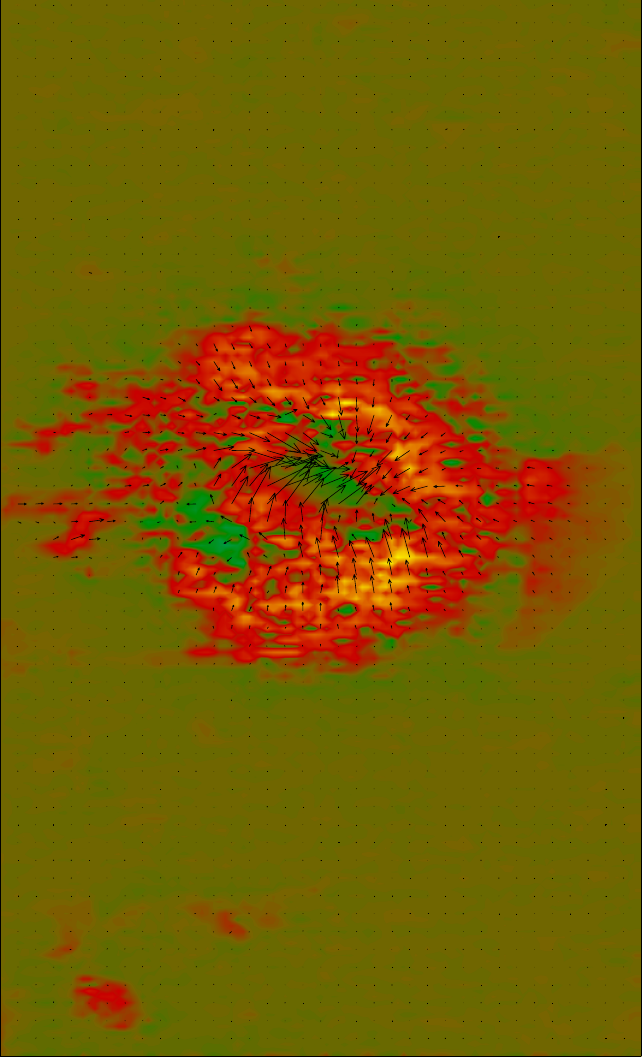}
\includegraphics[width=1.5cm]{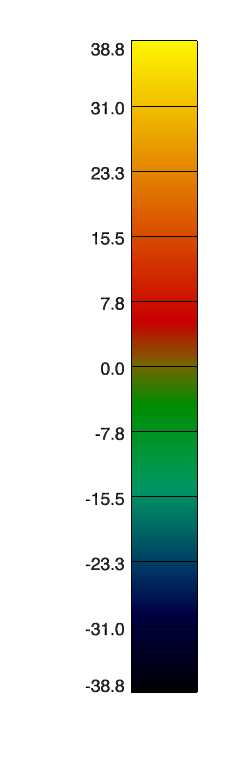}
\includegraphics[width=1.5cm]{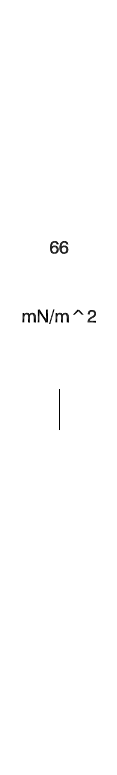}
\caption{West sunspot in NOAA 14142 observed on July 16, 2025: vector Lorentz force.}
\label{F-16July}
\end{figure}

\begin{figure}
\includegraphics[width=5.8cm]{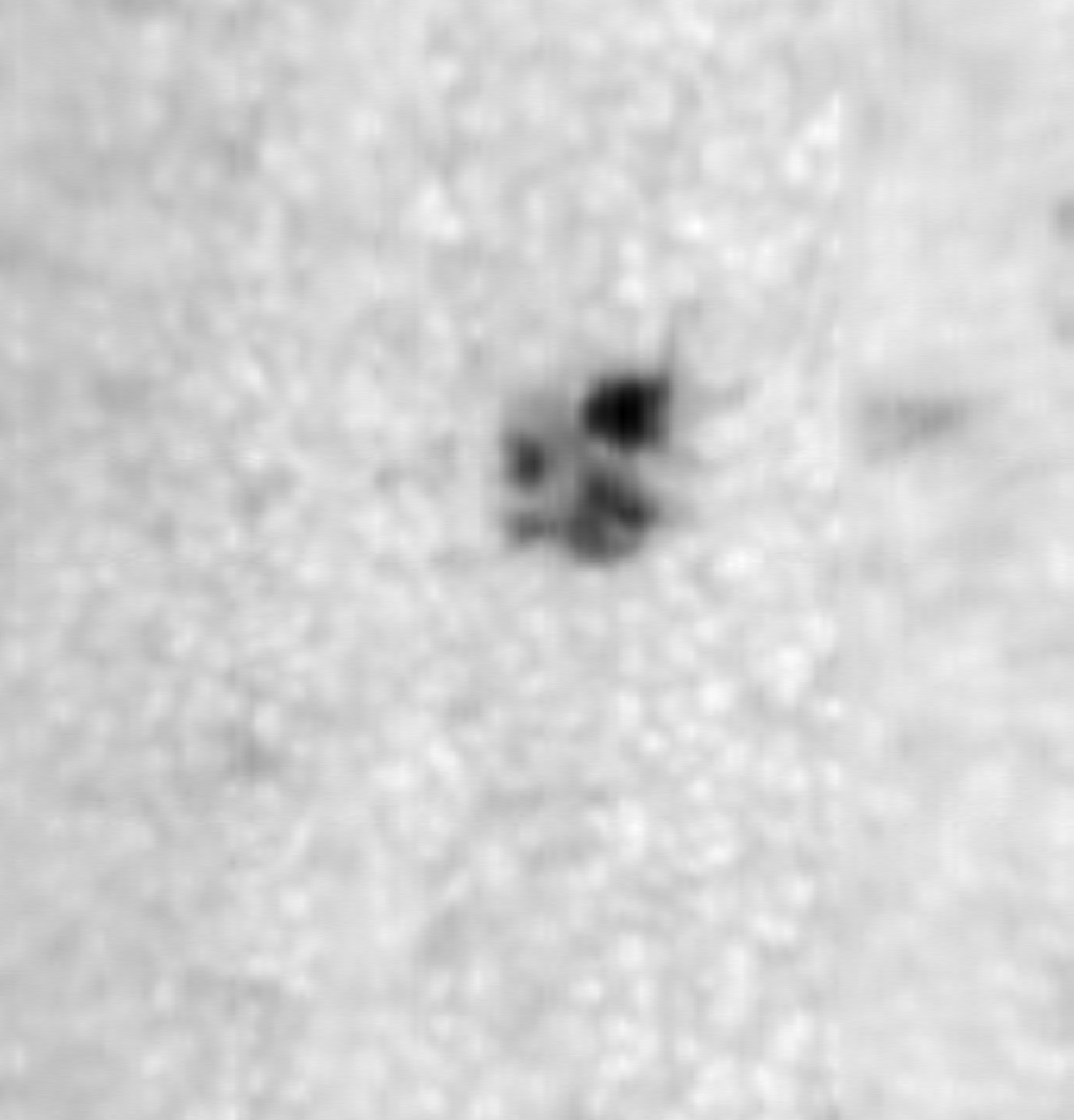}
\caption{Central region in NOAA 14142 observed on July 17, 2025.}
\label{I-17July}
\end{figure}

\begin{figure}
\includegraphics[width=5.8cm]{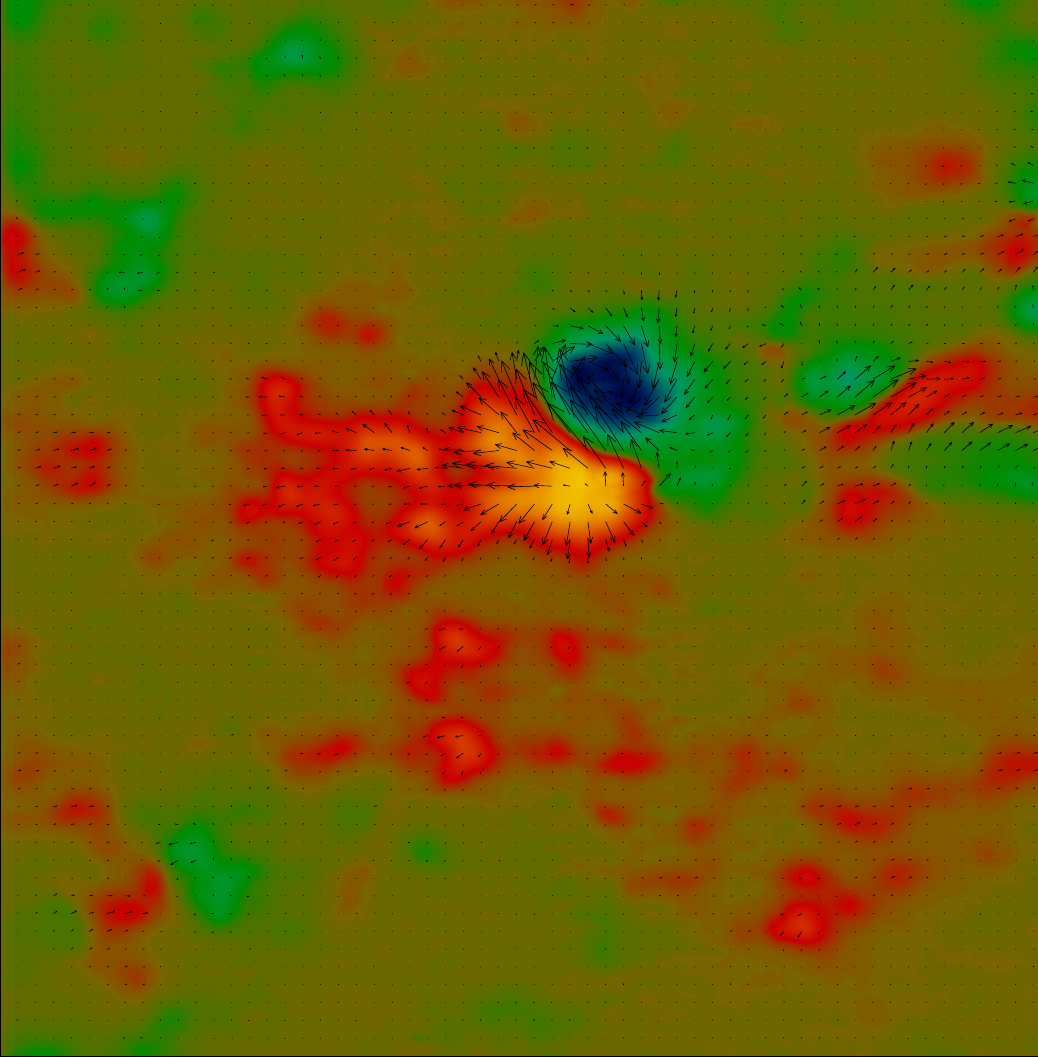}
\includegraphics[width=1.5cm]{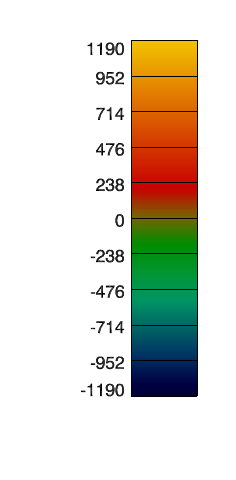}
\includegraphics[width=1.5cm]{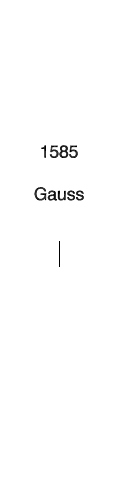}
\caption{Central region in NOAA 14142 observed on July 17, 2025: vector magnetic field.}
\label{B-17July}
\end{figure}

\begin{figure}
\includegraphics[width=5.8cm]{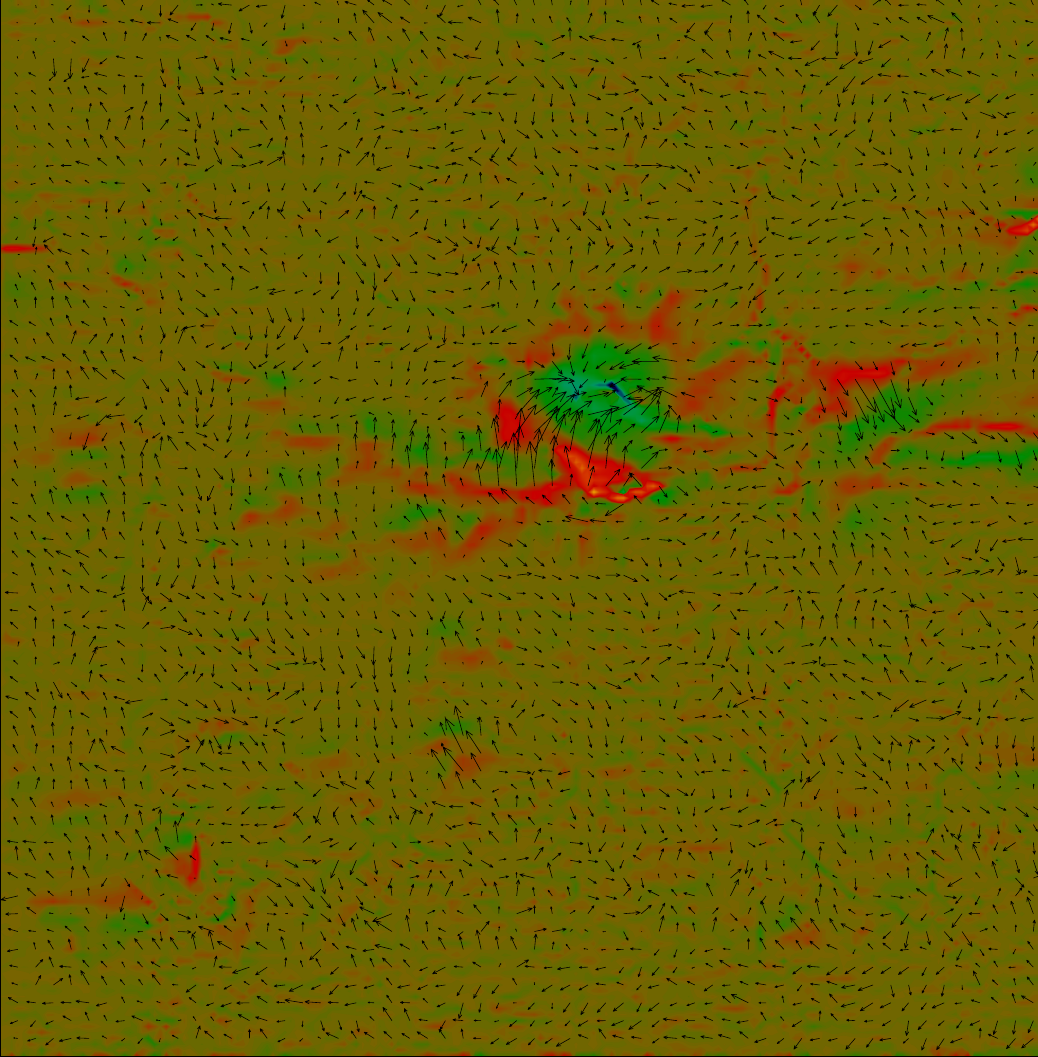}
\includegraphics[width=1.5cm]{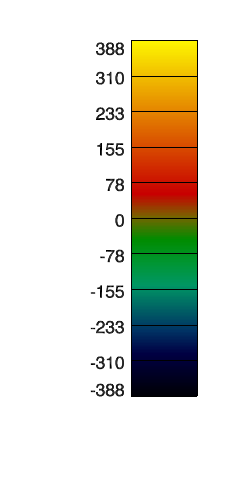}
\includegraphics[width=1.5cm]{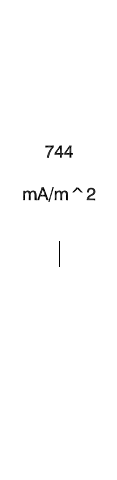}
\caption{Central region in NOAA 14142 observed on July 17, 2025: vector current density.}
\label{J-17July}
\end{figure}

\begin{figure}
\includegraphics[width=5.8cm]{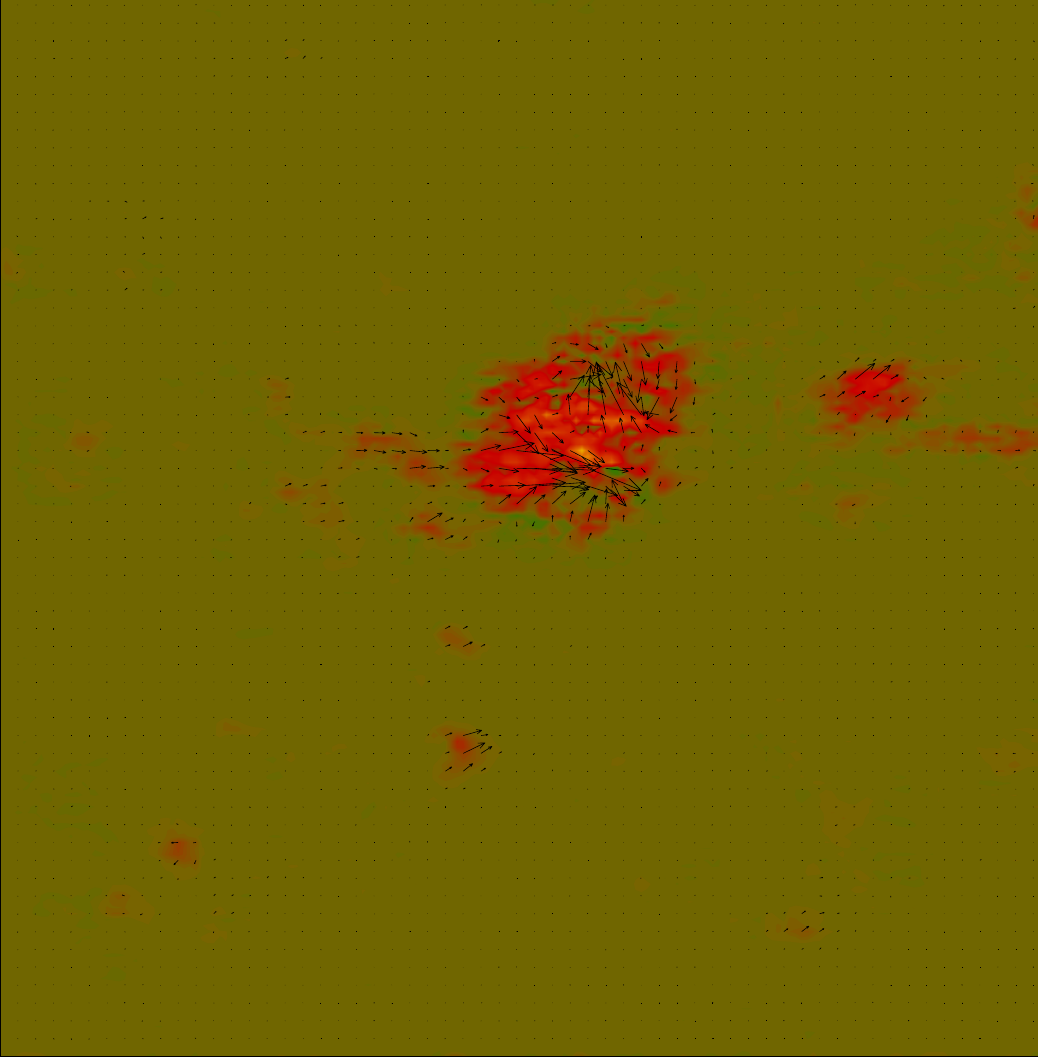}
\includegraphics[width=1.5cm]{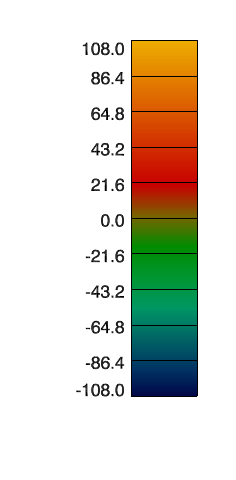}
\includegraphics[width=1.5cm]{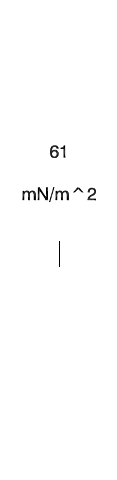}
\caption{Central region in NOAA 14142 observed on July 17, 2025: vector Lorentz force.}
\label{F-17July}
\end{figure}

We applied those conditions to new observations obtained in July 2025. We observed the active region NOAA 14142. The weather for image quality was not excellent. We measured polarisation in two lines: Fe I 6301.5 and 6302.5 \AA. These two lines are formed in the photosphere, but at different depths as observed by \citet{Faurobert-etal-09}. This provides the vertical variation of the magnetic field, which enables the ambiguity resolution and derivation of the vector current density as described in \citet{Bommier-25}. Then, the vector Lorentz force can be derived as final result. The inversion of polarimetric data into magnetic field data was performed by applying the Milne-Eddington inversion codes UNNOFIT (for normal Zeeman triplet lines like Fe I 6302.5 \AA) and UNNOFIT2 (for more complex lines like Fe I 6301.5 \AA), as described in \citet{Bommier-etal-07}. 

We observed an isolated spot (Fig. \ref{I-16July}) and two close sunspots (Fig. \ref{I-17July}). The magnetic field was found diverging from positive polarity sunspots (Figs. \ref{B-16July} and \ref{B-17July}) and converging towards center in the negative polarity sunspot (Fig. \ref{B-17July}). In those Figures, the vector magnetic field is plotted in solar coordinates: solar vertical and horizontal components. 

As in \citet{Bommier-25}, the vector density current, which is mainly horizontal, is found to be wrapping spots clockwise about a positive polarity spot (Figs. \ref{J-16July} and  \ref{J-17July}) and counterclockwise about the negative polarity spot (Fig. \ref{J-17July}). Also as in \citet{Bommier-25}, we observe a strong horizontal current component crossing the active region neutral line (Fig. \ref{J-17July}). Also as in \citet{Bommier-25}, the Lorentz force vector is systematically found centripetal with respect to the spot center, whatever the spot polarity is (see Figs. \ref{F-16July} and \ref{F-17July}). The Lorentz force thus maintains sunspots, which seems reasonable.

\section{Conclusion}

We hope to have shown with our test, that the TH\'EMIS telescope remains a calibration-free telescope, even after Adaptive Optics inclusion. The telescope is now able to provide vector magnetic field results, vector current density results, and vector Lorentz force results, which are in accordance with previous ones.

\begin{acknowledgements}
Thanks to Claude Le Men for the optics modifications.
\end{acknowledgements}

\bibliographystyle{aa}
\bibliography{bommierrefs}

\begin{thebibliography}{6}
\expandafter\ifx\csname natexlab\endcsname\relax\def\natexlab#1{#1}\fi

\bibitem[{{Bommier}(2025)}]{Bommier-25}
{Bommier}, V. 2025, \aap, 694, A40

\bibitem[{{Bommier} {et~al.}(2007){Bommier}, {Landi Degl'Innocenti},
  {Landolfi}, \& {Molodij}}]{Bommier-etal-07}
{Bommier}, V., {Landi Degl'Innocenti}, E., {Landolfi}, M., \& {Molodij}, G.
  2007, \aap, 464, 323

\bibitem[{{Bommier} \& {Rayrole}(2002)}]{Bommier-Rayrole-02}
{Bommier}, V. \& {Rayrole}, J. 2002, \aap, 381, 227

\bibitem[{{Faurobert} {et~al.}(2009){Faurobert}, {Aime}, {P{\'e}rini},
  {Uitenbroek}, {Grec}, {Arnaud}, \& {Ricort}}]{Faurobert-etal-09}
{Faurobert}, M., {Aime}, C., {P{\'e}rini}, C., {et~al.} 2009, \aap, 507, L29

\bibitem[{{Gelly} {et~al.}(2016){Gelly}, {Langlois}, {Moretto}, {Douet}, {Lopez
  Ariste}, {Tallon}, {Thi{\'e}baut}, {Geyskens}, {Lorgeoux}, {L{\'e}ger}, \&
  {Le Men}}]{Gelly-etal-16}
{Gelly}, B., {Langlois}, M., {Moretto}, G., {et~al.} 2016, in Society of
  Photo-Optical Instrumentation Engineers (SPIE) Conference Series, Vol. 9906,
  Ground-based and Airborne Telescopes VI, ed. H.~J. {Hall}, R.~{Gilmozzi}, \&
  H.~K. {Marshall}, 99065A

\bibitem[{{Schmieder} {et~al.}(2025){Schmieder}, {Bommier}, \&
  {Gelly}}]{Schmieder-etal-25}
{Schmieder}, B., {Bommier}, V., \& {Gelly}, B. 2025, Universe, 11, 153

\end{thebibliography}

\end{document}